\documentclass[10pt,a4paper]{article}

\newcommand{\Ap}{{A^\prime}}
\newcommand{\App}{{A^{\prime\prime}}}
\newcommand{\Bp}{{B^\prime}}
\newcommand{\Bpp}{{B^{\prime\prime}}}
\newcommand{\Cp}{{C^\prime}}
\newcommand{\Cpp}{{C^{\prime\prime}}}

\begin{document}

\title{Global SSS space-time models\\
{\normalsize nothing}}
\author{Ll. Bel\thanks{e-mail:  wtpbedil@lg.ehu.es}}

\maketitle

\begin{abstract}

We discuss Global Static Spherically Symmetric space-time models of mass $m$ with regular sources at the origin and asymptotically Minkowskian behavior at infinity; the interior model and the exterior one being matched at the radius $R$ of the source in the sense of Lichnerowicz. The global models depend in general on $R$ through a function $Q$ of $m$ and $R$.  Although $R$ would be an spurious parameter if the exterior model was considered alone, it becomes intrinsic for the global model. The physical implication is that $R$ as well as $m$ determine, at some order of approximation, the dynamics of orbiting objects  or viceversa that this dynamics puts conditions on the physical state of the source.
\end{abstract}

\section{Global SSS space-time models}

Using Weyl's like decomposition, and obvious notations, we consider the line-element:

\begin{equation}
\label{1}
ds^2=-A^2dt^2+A^{-2}d\bar{s}^2, \quad A=A(r)
\end{equation}
with:

\begin{equation}
\label{2}
d{\bar s}^2=B^2dr^2+BCr^2d\Omega^2, \quad B=B(r),\ C=C(r)
\end{equation}

{\it Space and Reference models}. We shall refer to this 3-dimensional metric as the space model, and to restrict it as well as the coordinates used we shall require the condition:

\begin{equation}
\label{3}
\Cp=\frac{2}{r}(B-C)
\end{equation}
Introducing the Euclidean reference metric:

\begin{equation}
\label{4}
d{\tilde s}^2=dr^2+r^2d\Omega^2,
\end{equation}
we can write the equation above as follows:

\begin{equation}
\label{5}
(\bar\Gamma^i_{jk}-\tilde\Gamma^i_{jk})g^{jk}=0, \quad i,j,k=1,2,3
\end{equation}
where $\bar\Gamma$ and $\tilde\Gamma$ are the connection symbols of (\ref{2}) and (\ref{4}) in which case this equation becomes a restriction on the model but remains true whatever space coordinates we use. In particular Cartesian coordinates of (\ref{4}) become harmonic coordinates of (\ref{2}), and therefore, as it is known, they are also harmonic coordinates of (\ref{1}).

Let $\bar B(\bar r)$ and $\bar C(\bar r)$ be a particular solution of Eq. (\ref{3}):

\begin{equation}
\label{8}
\frac{d\bar C}{d\bar r}=\frac{2}{\bar r}(\bar B-\bar C)
\end{equation}
and let us consider a coordinate transformation of the coordinate $r$:

\begin{equation}
\label{6}
r=r(\bar r)
\end{equation}
This will induce the following transformation of the metric potentials:

\begin{equation}
\label{7}
B=\bar B\left(\frac{dr}{d\bar r}\right)^{-1}, 
\quad C=\bar C\frac{\bar r^2}{r^2}\frac{dr}{d\bar r}
\end{equation}
As a direct calculation then proves $B(r)$ and $C(r)$ will satisfy Eq. (\ref{3}) if $r(\bar r)$ is a solution of:

\begin{equation}
\label{9}
\bar r^2\frac{d\bar C}{d\bar r}\frac{dr}{d\bar r}
+2\bar C\bar r\frac{dr}{d\bar r}+\bar C\bar r^2\frac{d^2 r}{d\bar r^2}-2\bar B r=0
\end{equation}
The above equation defines the restricted covariance of the model.

{\it Einstein's tensor}.- We write below the non identically zero components of the Einstein's tensor:

\begin{eqnarray}
\label{10}
S^0_0=\frac{2A\App}{B^2}-\frac{3\Ap^2}{B^2}+\frac{4A\Ap}{rBC}-\frac{A^2}{r^2BC}+\frac{A^2}{r^2C^2}
-\frac{3A^2\Bp}{rB^2C}-\frac{A^2\Bpp}{B^3}+\frac{5A^2\Bp^2}{4B^4}\\
\label{11}
S^1_1=-\frac{A^2\Bp}{rB^2C}+\frac{A^2}{r^2BC}+\frac{\Ap^2}{B^2}-\frac{A^2}{r^2C^2}-\frac{A^2\Bp^2}{4B^4}\\
\label{12}
S^2_2=S^3_3=\frac{3A^2\Bp^2}{4B^4}-\frac{\Ap^2}{B^2}-\frac{A^2}{r^2BC}-\frac{A^2\Bpp}{2B^3}-\frac{A^2\Bp}{rB^2C}
+\frac{A^2}{r^2C^2}
\end{eqnarray}
They have been somewhat simplified eliminating the derivatives of $C$ using (\ref{3}) and its derivative:

\begin{equation}
\label{13}
\Cp=\frac{2}{r}(B-C), \quad \Cpp=\frac{2}{r^2}(3C-3B+r\Bp)
\end{equation}

{\it The global model}.- The global models that we consider will depend on two essential parameters: the mass $m$ of the source and its radius $R$. From $r=R$ to $r\rightarrow\infty$ the model will be an asymptotic approximation to the Schwarzshild vacuum solution using appropriate coordinates satisfying (\ref{3}); and from $r=0$ to $r=R$ the model will be defined by regular series approximations. At $r=R$ the two models will be matched in the sense of Lichnerowicz, i.e., requiring the continuity of all the potentials and its derivatives. We do not solve Einstein's equations but we use the Einstein's tensor to define the density of the source and its two pressures: the radial pressure $P_r$ and the tangential one $P_t$ in general non equal:

\begin{equation}
\label{14}
\rho=S^0_0, \quad P_r=-S^1_1, \quad P_t=-S^2_2=-S^3_3
\end{equation}

\section{The exterior domain}

We start the description of the vacuum domain with Fock's line-element \cite{Fock} defined by the three potentials:

\begin{equation}
\label{15}
\bar A=\sqrt{\frac{\bar r-m}{\bar r+m}}, \ \bar B=1,  \ \bar C=1-\frac{m^2}{\bar r^2}
\end{equation}
This line element satisfies the harmonic condition (\ref{3}).

Substituting these functions in (\ref{9}) we obtain the equation that a function $r(\bar r)$ has to satisfy:

\begin{equation}
\label{16}
2{\bar r}\frac{dr}{d{\bar r}}+({\bar r}^2-m^2)\frac{d^2r}{d{\bar r}^2}-2r=0
\end{equation}
so that $B$ and $C$ defined by (\ref{7}) satisfy (\ref{3}). The general solution, taking into account the asymptotic condition $r\rightarrow\bar r$ at infinity can be written as follows:

\begin{equation}
\label{17}
r=\bar r+Q\left(1+\frac{\bar r}{2m}\ln\left(\frac{\bar r-m}{\bar r+m}\right)\right),
\end{equation}
where $Q$ is a constant that will latter choose to be a function of $m$ and $R$.

The first two terms of the asymptotic expansion of $r$ as a function of $\bar r$ are:

\begin{equation}
\label{18}
r=\bar r-\frac{1}{3}Q\frac{m^2}{{\bar r}^2};
\end{equation}
and conversely:

\begin{equation}
\label{19}
\bar r=r+\frac{1}{3}Q\frac{m^2}{r^2};
\end{equation}
Using both expansions and the corresponding expansions derived from (\ref{7}) a straightforward calculation yields the general form of the third order asymptotic expansions of the potentials $A,B,C$ as functions of $r$, satisfying the harmonic condition (\ref{3}) for the Schwarzschild vacuum solution:

\begin{eqnarray}
\label{20}
A=1-\frac{m}{r}+\frac12\frac{m^2}{r^2}-\frac12\frac{m^3}{r^3}\\
\label{21}
B=1-\frac23Q\frac{m^2}{r^3}\\
\label{22}
C=1-\frac{m^2}{r^2}+\frac43 Q\frac{m^2}{r^3}
\end{eqnarray}

\section{The interior domain}

To describe the source model we postulate to describe the potentials $A,B,C$ in the domain $r=0..R$, with $p=1,2,3$ in succession,  by the following double series:

\begin{eqnarray}
\label{23}
A^p_i=1+\sum_{n=0}^p\sum_{s=1}^pA_{ns}r^{2n}m^s,\\
\label{24}
B^p_i=1+\sum_{n=0}^p\sum_{s=1}^pB_{ns}r^{2n}m^s,\\
\label{25}
C^p_i=1+\sum_{n=0}^s\sum_{s=1}^pC_{ns}r^{2n}m^s
\end{eqnarray}
so that for example with $p=3$:

\begin{eqnarray}
\label{26}
A_i=1+m(A_{01}+A_{11}r^2)+m^2(A_{02}+A_{12}r^2+A_{22}r^4)\nonumber\\
+m^3(A_{03}+A_{13}r^2+A_{23}r^4+A_{33}r^6);
\end{eqnarray}
and similar expressions for $B$ and $C$. These expansions have been truncated at the order 3 in $m$ to match the order of approximation of the asymptotic expansions (\ref{20})-(\ref{22}) that are also polynomials on $m$. Notice also that the expressions above guarantee the regularity of the model at the origin:

\begin{equation}
\label{27}
\left(\frac{dA_i}{dr}\right)_{r=0}=0, \ \left(\frac{dB_i}{dr}\right)_{r=0}=0, \ \left(\frac{dC_i}{dr}\right)_{r=0}=0,
\end{equation}
as well as the symmetry $r\rightarrow -r$

To have a well defined global model, we have to implement now Eq. (\ref{3}) in a domain $0\leq r \leq R$, $R$ being the prescribed radius of the source. It is very simple to check that this leads to the following relationships:

\begin{equation}
\label{28}
B_{ns}=(n+1)C_{ns}, \ s=1,2,3 \ \ n\leq s
\end{equation}
that we shall always take into account.

\section{Matching the interior to the exterior}

To proceed to match the interior domain to the exterior one by equating equal powers of $m$  in the developments of the potentials evaluated at $r=R$ we choose to assume that the function $Q$ of $m$ that appears in (\ref{21}) and (\ref{22}) can be written:

\begin{equation}
\label{29}
Q=Q_0+mQ_1
\end{equation}
and therefore we rewrite them as follows:

\begin{equation}
\label{30}
A_e=1+\sum_{s=1}^3A_s\frac{m^s}{r^s},\ A_1=-1,\ A_2=\frac12, \ A_3=-\frac12
\end{equation}

\begin{equation}
\label{31}
B_e=1+\sum_{s=1}^3B_s\frac{m^s}{r^s},\ B_1=0, \ B_2=-\frac{2}{3}Q_0, \ B_3=-\frac23 Q_1
\end{equation}

\begin{equation}
\label{32}
C_e=1+\sum_{s=1}^3C_s\frac{m^s}{r^s},\ C_1=0, \ C_2=-1, \ C_3=-\frac43 Q_0;
\end{equation}
and introduce also the notations:

\begin{eqnarray}
\label{33}
A^{(p)}_i=1+\sum_{s=1}^p\sum_{n=0}^sA_{ns}r^{2n}m^s \\
\label{34}
B^{(p)}_i=1+\sum_{s=1}^p\sum_{n=0}^sB_{ns}r^{2n}m^s \\
\label{35}
C^{(p)}_i=1+\sum_{s=1}^p\sum_{n=0}^sC_{ns}r^{2n}m^s
\end{eqnarray}
Solving the matching problem order by order means solving one after the other the three systems ($p=1,2,3$) of linear equations :

\begin{equation}
\label{36}
(A^{(p)}_i)_{r=R}=(A^{(p)}_e)_{r=R}, \ \
\left(\frac{dA^{(p)}_i)}{dr}\right)_{r=R}=\left(\frac{dA^{(p)}_e)}{dr}\right)_{r=R}
\end{equation}

\begin{equation}
\label{37}
(B^{(p)}_i)_{r=R}=(B^{(p)}_e)_{r=R}, \ \
\left(\frac{dB^{(p)}_i)}{dr}\right)_{r=R}=\left(\frac{dB^{(p)}_e)}{dr}\right)_{r=R}
\end{equation}

\begin{equation}
\label{38}
(C^{(p)}_i)_{r=R}=(C^{(p)}_e)_{r=R}, \ \
\left(\frac{dC^{(p)}_i)}{dr}\right)_{r=R}=\left(\frac{dC^{(p)}_e)}{dr}\right)_{r=R}
\end{equation}

We list below the restrictions that each of these systems imposes:

$p=1$:

\begin{equation}
A_{00}=1, \quad
A_{01}=-\frac{3}{2R},\quad A_{11}=\frac{1}{2R^3}
\end{equation}

\begin{equation}
B_{00}=1, \quad
B_{01}=0,\quad B_{11}=0
\end{equation}

\begin{equation}
C_{00}=1, \quad
C_{01}=0,\quad C_{11}=0
\end{equation}

$p=2$:

\begin{equation}
A_{02}=A_{22}R^4+\frac{1}{R^2},\quad A_{12}=-2A_{22}R^2-\frac{1}{2R^4}
\end{equation}

\begin{equation}
B_{02}=\frac{1}{6}\frac{-18R+35Q_0}{R^3}, \quad B_{12}=-2\frac{-3R+7Q_0}{R^5}, \quad B_{22}=\frac{3}{2}\frac{-2R+15Q_0}{R^7}
\end{equation}

\begin{eqnarray}
C_{02}=\frac{1}{6}\frac{-18R+35Q_0}{R^3}, \\
C_{12} =-\frac{-3R+7Q_0}{R^5}, \quad C_{22}= \frac{1}{2}\frac{-2R+5Q_0}{R^7}
\end{eqnarray}

$p=3$

\begin{eqnarray}
&& A_{03}=A_{23}R^4+2A_{33}R^6-\frac{5}{4R^3}+\frac{Q_0}{R^4}, \\[2ex]
&& A_{13}=-2A_{23}R^2-3A_{33}R^4+\frac{3}{4R^5}-\frac23\frac{Q_0}{R^6}
\end{eqnarray}

\begin{eqnarray}
&& B_{03}=-\frac{1}{6}\frac{6C_{33}R^9-35Q_1}{R^3}, \quad
B_{13}=2\frac{3C_{33}R^9-7Q_1}{R^5}, \nonumber \\[2ex]
&& \hspace {35mm} B_{23}=-\frac32\frac{6C_{33}R^9-5Q_1}{R^7}
\end{eqnarray}

\begin{eqnarray}
&& C_{03}=-\frac{1}{6}\frac{6C_{33}R^9-35Q_1}{R^3}, \quad
C_{13}=2\frac{3C_{33}R^9-7Q_1}{R^5}, \nonumber \\[2ex]
&& \hspace {35mm}C_{23}=-\frac12\frac{6C_{33}R^9-5Q_1}{R^7}
\end{eqnarray}

At this point the coefficients $A_{22},A_{23},A_{33},C_{33}$ and
$Q_0,Q_1$ remain as free parameters. Selecting a particular value of the density $\rho_0$ at the origin and a particular value of the tension $(P_r)_0$ which is always equal to $(P_t)_0$ reduces the number of free parameters, not including $m$ and $R$, to four.

\section{Particular models}

{\it Perfect fluids}.-The physical interpretation of the approximate solutions thus obtained rests on the values of the density and the internal tensions that can be derived from them using the Einstein´s equations (\ref{10})-(\ref{12}). Of particular interest will be those for which the two tensions $P_r$ and $P_t$ coincide and we may think of the interior as being a perfect fluid. By direct substitution of the values of $A_i,B_i,C_i$ obtained above it follows from (\ref{11}) and (\ref{12}) that for this to be the case we have to have:

\begin{eqnarray}
-\frac{12}{R^6}+35\frac{Q_0}{R^7}=0 \\
-\frac{7(6C_{33}R^9-5Q_1)}{R^7}+\frac{38}{R^7}-\frac{16A_{22}}{R}-\frac{105Q_0}{R^8}=0 \\
54C_{33}+\frac{16A_{22}}{R^3}+\frac{35Q_0}{R^{10}}-\frac{14}{R^9}=0
\end{eqnarray}
wherefrom we get:

\begin{equation}
\label{52}
A_{22}=\frac{1}{32}\frac{315Q_1+4}{R^6}, \ C_{33}=-\frac{Q_1}{R^9}, \ Q_0=\frac{12}{35}R
\end{equation}
At this point $A_{23}, A_{33}$ and $Q_1$ remain as free parameters.

It is noteworthy to remark that Fock's model with $Q=0$ can not be matched with an interior domain described by a perfect fluid.

{\it Constant density}.- Although one of the interests of this paper is to produce approximate but rather general global models it might be occasionally useful to consider also models with constant density. Thus, using the Einstein equation (\ref10{}) to calculate the density and requiring it to be constant yields the following equations:

\begin{eqnarray}
40A_{22}+\frac{70}{R^6}-\frac{175Q_0}{R^7}=0\\
40A_{23}+\frac{35(6C_{33}R^9-5Q_1)}{R^7}-\frac{315}{R^7}-\frac{60A_{22}}{R}-\frac{77Q_0}{R^8}=0 \\
84A_{33}-189C_{33}+\frac{2A_{22}}{R^3}-\frac{245Q_0}{R^{10}}-\frac{98}{R^9}=0
\end{eqnarray}
wherefrom we get:

\begin{eqnarray}
&& A_{22}=\frac{7}{8}\frac{-2R+5Q_0}{R^7}, \quad A_{23}=-\frac{7}{24}\frac{8R^{10}A_{33}-9R+21Q_0-15RQ_1}{R^8}, \nonumber \\[2ex]
\label{56}
&& \hspace {35mm} C_{33}=\frac{1}{36}\frac{16R^{10}A_{33}+18R-45Q_0}{R^{10}}
\end{eqnarray}

{\it Perfect fluid with constant density}.-Assuming now that the source is a perfect fluid and that the density is constant, using therefore both (\ref{52}) and (\ref{56}) we obtain:

\begin{eqnarray}
Q_0=\frac{12}{35}R, \ Q_1=-\frac{4}{105},\\
A_{22}=-\frac{1}{4R^6}, \ A_{33}=\frac{5}{56R^9}\\
C_{33}=\frac{1}{9R^{9}}
\end{eqnarray}
For this case the free parameters are both $m$ and $R$.

\section{Discussion}

This paper provides the interested reader with a fairly large class of global SSS space-time approximate models with a variety of interior domains leading to exterior ones that depend on $m$ but also on $R$. It is worthy to insist in saying that the restricted covariance of the models applied to the exterior domain would always allow to get rid of $R$ but this is not true for the global model. A few papers \cite{Liu}-\cite{CMMR} pointed already in this direction but this fact has apparently remained unnoticed.

It follows from this that, beyond the second order approximation, the relativistic corrections might be  not only quantitatively important but also qualitatively. To be more precise about the physical implications of our models, let us consider the dynamics of a massif test particle orbiting the plane $\theta=\pi/2$ in the exterior domain. $\epsilon$  and $\mu$ being respectively the relativistic energy and angular momentum per unit mass we obtain the orbital equation:

\begin{equation}
\left(\frac{d\phi}{du}\right)^2=F_0+F_1u+F_2u^2+F_3u^3+F_4u^4, \ u=\frac{1}{r}
\end{equation}
where:

\begin{eqnarray}
\label{62}
F_0=\frac{\epsilon^2-1}{\mu^2}, \ F_1=\frac{2m(2\epsilon^2-1)}{\mu^2},
\ \ F_2=-1+\frac{6m^2\epsilon^2}{\mu^2}, \\
F_3=\frac{8}{3}\frac{m^2(\epsilon^2-1)Q_0}{\mu^2}+\frac{8}{3}\frac{m^3(\epsilon^2-1)Q_1}{\mu^2}+
\frac{2m^3(2\epsilon^2+1)}{\mu^2}\\
F_4=m^2+\frac{14}{3}\frac{m^3(2\epsilon^2-1)Q0}{\mu^2}
\end{eqnarray}
and if instead of a massive particle we consider a light ray whose minimum value of $r$ is $b$ then, considering the limits:

\begin{equation}
\epsilon\rightarrow 0, \ \mu\rightarrow 0, \ \frac{\epsilon}{\mu}\rightarrow \frac{1}{b}
\end{equation}
the preceding coefficients become:

\begin{eqnarray}
\label{66}
F_0=\frac{1}{b^2}, \ F_1=\frac{4m}{b^2},
\ \ F_2=-1+\frac{6m^2}{b^2}, \\
F_3=\frac{8}{3}\frac{m^2Q_0}{b^2}+\frac{8}{3}\frac{m^3Q_1}{b^2}+
\frac{4m^3}{b^2}\\
F_4=m^2+\frac{28}{3}\frac{m^3Q0}{b^2}
\end{eqnarray}

To derive the two emblematic results that confirmed  Einstein´s theory only the three first terms of either (\ref{62}) or (\ref{66}) are needed and the value of $Q_0$ will become important only when higher order precision astronomical observations will become available.

\section*{Acknowledgments}

I gratefully acknowledge the careful reading of this manuscript by J. Mart\'{\i}n
and the useful comments he made to me.

\end{document}